\documentclass[aps,10pt,prb,twocolumn,superscriptaddress,showpacs,floats,floatfix,preprintnumbers,amsmath,amssymb]{revtex4-1}
\usepackage[T1]{fontenc}
\usepackage[USenglish,american]{babel}
\usepackage{verbatim}
\usepackage{amsmath}
\usepackage{amssymb}
\usepackage{wasysym}
\usepackage{graphicx}
\usepackage{tabularx}
\usepackage{xcolor}
\usepackage{bm}          
\usepackage{pifont} 
\usepackage{dcolumn}
\usepackage{float}
\usepackage{natbib}

\usepackage{subfigure,textcomp,epsfig,siunitx}
\usepackage{array}
\usepackage{longtable}
\usepackage{bm}
\usepackage{booktabs}

\graphicspath{{plots/}}

\newcommand{\vol}{volume}
\newcommand{\density}{$\rho_V$}
\newcommand{\nH}{$n_H$}
\newcommand{\dAH}{d$_{\textrm{AH}}$}
\newcommand{\dHH}{d$_{\textrm{HH}}$}
\newcommand{\spg}{$SPG$}

\newcommand{\dH}{$\Delta$H}

\newcommand{\TDOS}{TDOS}
\newcommand{\ADOS}{ADOS}
\newcommand{\HDOS}{HDOS}

\newcommand{\rhoA}{$\rho_A$}
\newcommand{\rhoH}{$\rho_H$}

\newcommand{\SHYDRA}{\texttt{Superhydra}}

\newcommand{\BMq}{BM$_q$}

\newcommand{\wgmin}{$\omega_{\Gamma}^{min}$}
\newcommand{\wgmax}{$\omega_{\Gamma}^{max}$}
\newcommand{\wgavg}{$\omega_{\Gamma}^{avg}$}

\newcommand{\elph}{$e-ph$}

\newcommand{\wLOG}{$\omega_{log}$}

\newcommand{\tc}{{$T_c$}}

\newcommand{\GrazPhys}{Institute of Theoretical and Computational Physics,
Graz University of Technology, NAWI Graz, 8010 Graz, Austria}
\newcommand{\UniRoma}{Dipartimento di Fisica, Universit\`a di Roma La Sapienza, Piazzale Aldo Moro 5, I-00185 Roma, Italy}
\newcommand{\CREF}{Centro Ricerche Enrico Fermi, Via Panisperna 89 A, 00184 Rome, Italy}

\begin{document}
\title{Mapping Superconductivity in High-Pressure Hydrides: The \SHYDRA{} Project.}

\author{Santanu Saha} \affiliation{\GrazPhys}
\email{santanu.saha@tugraz.at}
\author{Simone Di Cataldo} \affiliation{\GrazPhys}\affiliation{\UniRoma}
\author{Federico Giannessi} \affiliation{\UniRoma}
\author{Alessio Cucciari} \affiliation{\UniRoma}
\author{Wolfgang von der Linden} \affiliation{\GrazPhys}
\author{Lilia Boeri} \affiliation{\UniRoma}\affiliation{\CREF}

\date{\today}

\begin{abstract}
The discovery of high-\tc{} conventional superconductivity in high-pressure hydrides has helped
establish computational methods as a formidable tool to guide material discoveries in a field traditionally
dominated by serendipitous experimental search.
This paves the way to an ever-increasing use of data-driven approaches to the study and design of superconductors.
In this work, we propose a new method to generate meaningful datasets of superconductors, based on element substitution into a small
set of representative structural templates, generated by crystal structure prediction methods (MultiTemplate-HighThroughput approach).
Our approach realizes an optimal compromise between structural variety and computational efficiency, and can be easily generalized to
other elements and compositions.
As a first application, we apply it to binary hydrides at high pressure, realizing a database of 880 hypothetical
structures, characterized with a set of electronic, vibrational and chemical descriptors. 139 structures of our \SHYDRA{}
Database are superconducting according to the McMillan-Allen-Dynes approximation. Studying the distribution
of \tc{} and other properties across the database with advanced statistical and visualization techniques, we are able
to obtain comprehensive material maps of the phase space of binary hydrides.
The \SHYDRA{} database can be thought as a first step of a generalized effort to map conventional superconductivity.  
\end{abstract}

\maketitle

\begin{figure*}[htb]
\centering
\includegraphics[width=1.8\columnwidth,angle=0]{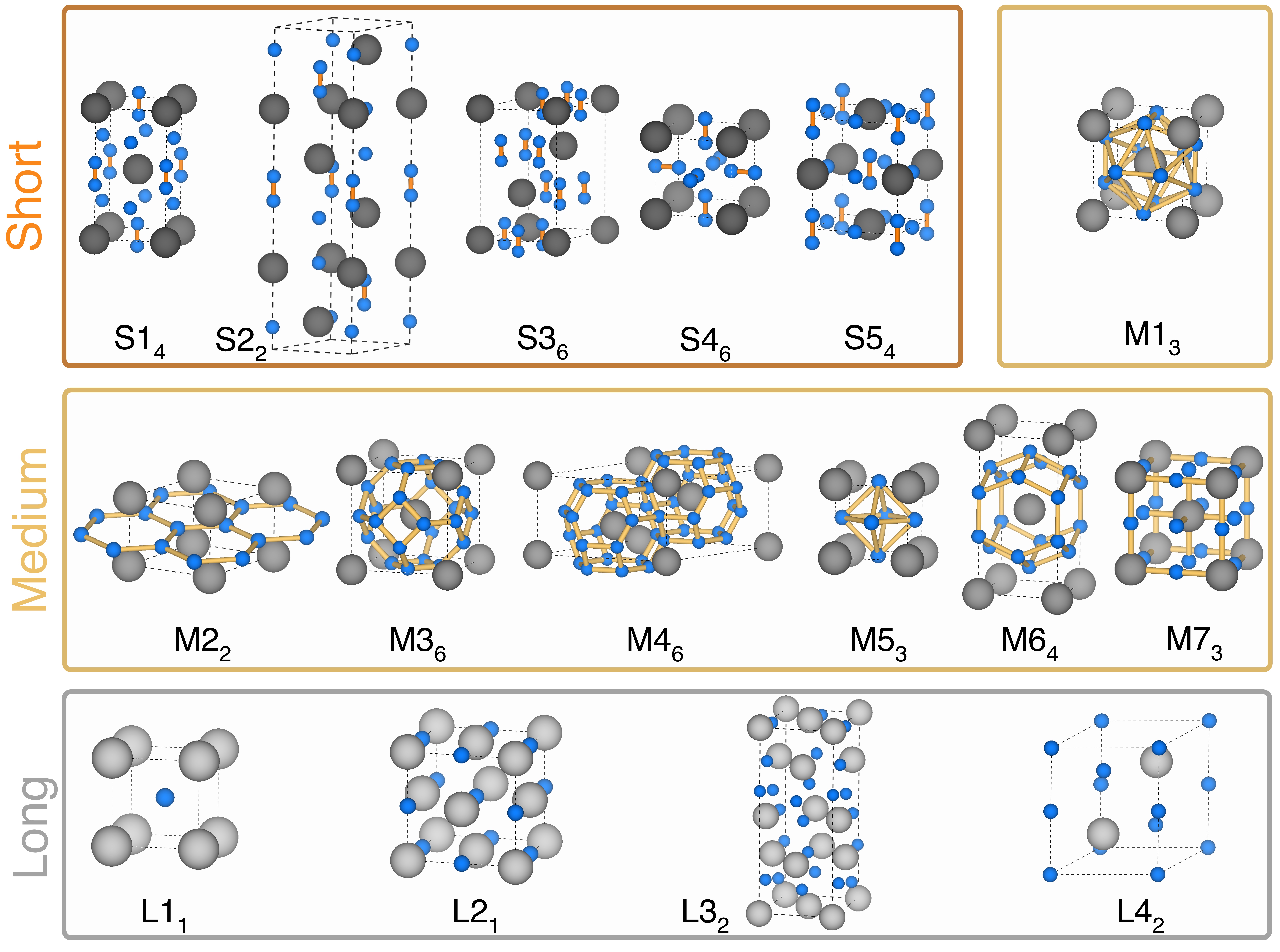}
\caption{
Template structures used for multi-template-high-throughput data generation, classified by H-H bond distance into short (S), medium (M), and long (L). The label $XN_{n_H}$ indicates the H-H distance type ($X$), the structure number ($N$), and the number of hydrogen atoms in the structure (subscript $n_H$)
Details of the crystal structures are given in the SI.
}
\label{fig:TEMPLATES}
\end{figure*}

\section{Introduction}\label{Sec:Intro}

Methods for {\em ab-initio}
prediction of superconductors led to an impressive acceleration in material discoveries,
following the breakthrough report of high-\tc{} superconductivity in high-pressure superhydrides\cite{duan2014pressure,drozdov2015conventional}.

After reaching the symbolical threshold of room-temperature\cite{snider2020room, drodzov2019lah10,somayazulu2019lah10},
the research focus is rapidly shifting to the even more ambitious goal of conventional superconductivity at ambient conditions \cite{dicataldo2021labh8,lucrezi2021basih8,dicataldo2021laxh,saha2020high}.
Meeting this challenge will require new methods which are able to explore a potentially huge material space,
much broader than that of simple binary hydrides;\cite{flores2020perspective,boeri20212021}
high-throughput (HT) screening and artificial intelligence techniques will likely play a prominent role in this search
 \cite{curtarolo2013high,hautier2010finding,glawe2016optimal,schmidt2018predicting,griesemer2021high,wang2021predicting, shipley2021high, semenok2020distribution}.
%

High-pressure superhydrides  represent an ideal benchmark for 
such data-driven approaches, since first-principles calculations can provide accurate estimates for \tc{}
and overcome the limitations of scarcity and inhomogeneity of data associated with other classes of superconductors.
Indeed, a few works in literature have attempted to 
study the distribution of \tc{}, in binary hydrides applying empirical classification schemes to data previously published in literature\cite{zurek2017alkalihydrides,semenok2020distribution,belli2021strong},
or randomly-generated structures\cite{shipley2021high,saha2020high}. 
Both approaches present obvious intrinsic drawbacks: while datasets based on literature 
are usually limited to near-ground-state structures, losing precious information on other possible chemical environments,
randomly-generated sets tend to contain many unphysical or unrepresentative structures \cite{rhorhofer2021features}.

In this paper we propose an alternative strategy for the exploration of the phase space of superconductors,
based on the high-throughput substitution of the elements of the periodic table in a small set of representative structural templates
for different compositions -- MultiTemplate-HighThroughput approach (MTHT) -- which 
is explicitly designed to realize the best possible compromise between variety, 
reproducibility, and efficiency. By utilizing the same set of high-symmetric, representative
templates for all elements of the periodic table, we reduce the computational time devoted
to structural search and focus our resources on computing the \tc{}, for a wider spectrum of structures,
regardless of their thermodynamic stability. 

As a first application, the MTHT approach is applied to map the phase space of binary hydrides formed by the first 55 elements of the periodic table (H and He are excluded).
Our \SHYDRA{} database, available as a single CSV file in the SM \footnote{The Supplemental Material is available at..}, 
comprises a total of $880$ distinct binary hydride structures, of which $139$ are superconducting and $27$ are high-\tc{};
all structures are characterized by a set of relevant structural, electronic, vibrational and superconducting properties,
computed at the same level of accuracy. 
Using advanced statistical methods to analyse our database, we obtain a set of
{\em material maps} depicting the distribution of physical properties across different families of binary hydrides, 
and identify general trends and correlations with thermodynamical stability and superconductivity.

This paper is structured as follows: in the first section we describe the
 construction of the \SHYDRA{} database{}, i.e. the generation of the 
 templates, the choice of the relevant features, and the classification
 of the hypothetical binary hydride structures into broad families, emerging from an advanced statistical analysis.
In the next section, we discuss the main features of the binary hydrides material space,
using material maps of the most relevant quantities.
The main computational details are summarized in the methods section; the SM
contains detailed information on the database generation methodology, additional descriptions of the database, the full set of material maps,
as well as additional information on the data analysis methods employed.

\section{Construction of the \SHYDRA{} Database}\label{Sec:ConstrDB}
The \SHYDRA{}  database was generated substituting the first 55 elements of the periodic table following hydrogen and helium (Li-La) 
into sixteen binary hydride templates AH$_n$, obtained with fixed-compositions evolutionary crystal structure searches for selected elements. 
Details on the generation procedure are available in the SM.

The resulting structures, relaxed at a common pressure of $200$ GPa, form a database of $55\times16=880$
hypothetical binary hydrides, for which we computed various electronic,
vibrational and thermodynamic properties, listed below (Database 1, or DB1 in 
the following). For the 139 metallic and dynamically stable
structures, we computed electron-phonon properties in the linear response approximation, and
estimated the \tc{} using the McMillan-Allen-Dynes formula,
assuming a constant $\mu^\ast=0.1$ \cite{PhysRev.167.331}. These compounds form our database of superconductors (DB2).
27 of these compounds are high-\tc, according to the conventional definition employed throughout
the manuscript, where the threshold  for high-\tc{} is set to the N$_2$ boiling point (77 $K$).


\subsection{Structural Templates}
The most critical issue in the construction of the \SHYDRA{} database was the choice of a
set of structural templates satisfying a few crucial requirements:
($i$) being sufficiently {\em representative}, i.e. the structural templates should
cover the variety of bonding environments encountered in known binary hydrides;
($ii$) at the same time,  the number of
templates should {\em not be too large} and ($iii$)  the unit cells of the 
individual templates should be small and possibly symmetric, to allow efficient
high-throughput calculations of electronic and vibrational properties. 
 Finally, ($iv$) the generation procedure should be {\em easily generalizable} to 
other systems and chemical compositions.

To meet these requirements, we proceeded as follows: we first identified a relatively small sets of
elements and compositions representative of different classes of binary hydrides:
low-\tc{} metallic hydrides; covalent hydrides like SH$_3$ and 
SeH$_3$ \cite{duan2014pressure, floreslivas2017seh3}; systems that form H$_2$-rich structures like TeH$_4$ and LiH$_{6}$ \cite{zhang2006structural, zurek2009lithiumhydrides}; cage-like structures, as in sodalite-clathrate hydrides like YH$_6$ \cite{heil2019superconductivity, belli2021strong,flores2020perspective}.
The elements were chosen to exhibit different chemical properties and high pressure behaviours. 
As for compositions, we restricted our search to compounds with chemical formulas AH$_n$, with $n \leq 6$, to ensure a relatively small unit cell
and a fast computation of the physical properties. 
With this choice, we consciounsly eliminated from the \SHYDRA{} database a large number of known high-\tc{} hydrides, such as LaH$_{10}$\cite{drodzov2019lah10} 
and YH$_{9}$\cite{kong2021yh6},
under the assumption that the main features of high-$n$ clathrates-like hydrides are well represented in the database by clathrates with $n=6$.

Once elements and compositions were decided, our generation procedure continued as follows:
first, we ran several unconstrained fixed-composition evolutionary
search runs for binary hydrides AH$_n$ of
selected elements (A = Li, Na, Ca, Cl, Y, Fe, S, C, Al)
and compositions ($n=1,2,3,4,6$) (see SM for details).
At the end of these structural searches, we retained
the 10 lowest enthalpy structures for each element and the structures with high symmetry (space group above 75), and filtered out all the rest.
After checking for duplicates, the structures were divided into three classes, based on the smallest H-H distance (\dHH): 
short (\dHH~$< 1$~\AA); medium ($1$~\AA~$\leq$~\dHH$~\leq 2$~\AA); long (\dHH~$> 2$~\AA).

The final set of sixteen templates was obtained retaining mostly cubic
structures (7), a few hexagonal (2), trigonal (3), tetragonal (3) ones, and one
orthorhombic, 
imposing that the set should contain at least two templates for each composition, and requiring that the templates would be balanced with respect to short, medium and long \dHH. 
Structural details for each template are provided in the SM. 

The sixteen templates are displayed in Fig.~\ref{fig:TEMPLATES}. In the following, they will be represented by an alphanumeric id of the form $XN_{n_H}$ where $X$ stands for short(S), medium(M) or long(L) type based on the smallest \dHH; $N$ represents the serial number and the subscript $n_H$, the number of H atoms in its chemical formula. Hence, e.g. M4$_6$ would represent the 4$^{th}$ structure of medium type with composition AH$_6$. 
Among the templates we find well-known crystal structure types. 
In particular, the M7$_3$ and the M3$_6$ structures correspond to the high-\tc{} 
SH$_3$ and AH$_6$ (A = Mg, Ca, Y) templates \cite{duan2014pressure, drozdov2015conventional, heil2015seh3sh3, feng2015mgh6, peng2017superconductivity, kong2021yh6}, while M1$_3$ and M5$_3$ correspond to the high-pressure phases of AlH$_3$ and FeH$_3$ \cite{pickard2007alh3, pepin2014FeH, heil2018absence}, respectively. L1$_1$ and L2$_1$ coincide with the CsCl and NaCl structures, and S5$_4$ with the predicted LiH$_6$ structure \cite{zurek2009lithiumhydrides}.

\subsection{Physical Properties - Features}
Here, we give a short description of the features grouped in different
categories; a more detailed description and definition of each feature is given in the
SM.
\begin{enumerate}
    \item Geometry - space group of the primitive unit cell (\spg); number of hydrogen atoms per formula unit (\nH); Volume(\vol);  Density (\density); Structural Template ($XN_{n_H}$); nearest-neighbour distance between H-H (\dHH); and between hydrogen and the guest atom A (\dAH). The volume and density is in \AA$^3$/atom and gm/cm$^3$ respectively. The inter-atomic distances are in \AA . 
    
    \item Thermodynamics - Formation enthalpy per atom calculated at 200 GPa (\dH{} in eV/atom). As reference for the pure elements, we employed structures from literature when available, or performed a dedicated evolutionary structural search with USPEX \cite{glass2006uspex}.
    
    \item Electronic Structure - Total Density of States (\TDOS), Partial Density of States (PDOS) on atom A/H (\ADOS; \HDOS) in states/eV/atom, Band gap (BG in eV); STATE (Metal or Insulator).
    
    \item Charge - Results of the Bader charge analysis on the unit cell; partial charge on atom A (\rhoA); partial charge on atom H (\rhoH), in a.u. - average, maximum, minimum value, standard deviation (avg, max, min, dev).

    \item Elastic - The elastic properties of the system are represented through
    quasi Bulk modulus estimated at 200 GPa (\BMq), in GPa.
    The definition of the \BMq~is provided in the SM.
    
    \item Phonon - Maximum, minimum and average phonon frequency at the $\Gamma$ point (\wgmin, \wgmax, \wgavg) in meV.
    
    \item Electron-phonon (\elph) - These properties have been evaluated for metallic dynamically stable structures only. 
    We report the logarithmic-averaged phonon frequency (\wLOG) in $K$, the total electron-phonon coupling constant ($\lambda$) and
    superconducting critical temperature (\tc{}) estimated from 
    McMillan-Allen-Dynes formula ($\mu^\ast=0.1$).
\end{enumerate}

\subsection{Statistical Analysis} 
Several empirical classification schemes have been proposed for binary hydrides,
based on structural or chemical properties, such as interatomic distances, guest atom charge
or atomic radius, electronegativity, electronic charge localization, etc. \cite{shipley2021high, belli2021strong, semenok2020distribution, stanev2018mltc}.

For example, it was early recognized that in order for a superhydride to exhibit high-\tc{} superconductivity, 
the interatomic H-H distance \dHH~should lie in a sweet spot between 0.9 and 1.35 \AA \cite{belli2021strong}, and
the hydrogen sublattice should be negatively charged \cite{wang2012cah}. Other authors recognized that high-\tc~superhydrides can be divided in two classes (covalent and clathrate hydrides), depending on the existence of a covalent bond between the A--H or a weakened H--H bond,~\cite{flores2020perspective} 
while Belli et al. divided the known superhydrides in six classes (molecular, covalent, weak H interaction, electride, isolated and ionic), 
using an inspection of interatomic distances and electronic localization function profiles \cite{belli2021strong}.

The \SHYDRA{} database is large and homogeneous enough to attempt an unbiased analysis of the 
properties distributions across different elements and structures, using  machine learning techniques.
The aim of this analysis is not to obtain accurate classifications or \tc{} prediction, but
rather to guide our study of the material maps letting correlations and trend emerge from the data \cite{pettifor1986structures}.

We first attempt to see whether the full database can be naturally divided into families, using all calculated features, except for  \elph~properties.
Unsupervised learning algorithms are designed specifically for the task of separating the data into families related to similarity of some sort, with no assumptions on the physical nature of this similarity. 

The $k$-means clustering algorithm, applied on to the whole \SHYDRA{} database (DB1), leads to the most physically-meaningful clusterization when $k = 3$ is employed \footnote{The choice of $k=3$ or $k=6$ is supported by a hierarchical clustering analysis (See Fig. 10 in SM).}. We name the three clusters thus identified H$_2$, TMs, and HTC,  based on the type of hydrides they comprise. The H$_2$ cluster contains primarily structures  with short H-H bonds (\dHH~$< 0.8$ ~\AA), i.e. molecular H$_2$ hydrides, 
covering all possible atomic numbers; the TMs (Transition Metal) cluster contains hydrides of the first-row $p$ elements and transition metals, covering all values of \dHH~larger than 1.0 \AA; the high-\tc{} (HTC) cluster comprises hydrides of the remaining elements, again with \dHH~$> 0.8$~\AA. 
The HTC cluster contains almost all of the high-\tc{} structures, hence its name.

This simple division into clusters is also rather efficient in
filtering out superconductors: out of the 27 hydrides with \tc{}
higher than 77 $K$,
85\% (23) belong to the HTC cluster; 4\% (1) to H$_2$; and 11\% (3) to TMs. The outlying high-\tc{} structures are S3$_6$-NaH$_6$ in the H$_2$ cluster, and S3$_{6}$-MoH$_6$, M3$_6$-NbH$_6$, and M4$_6$-NbH$_6$ in TMs. Known high-$T_c$ phases, such as SH$_3$, NaH$_6$ and YH$_6$ all belong to the HTC cluster.

\begin{figure}[!htb]
\centering
\includegraphics[width=0.95\columnwidth,angle=0]{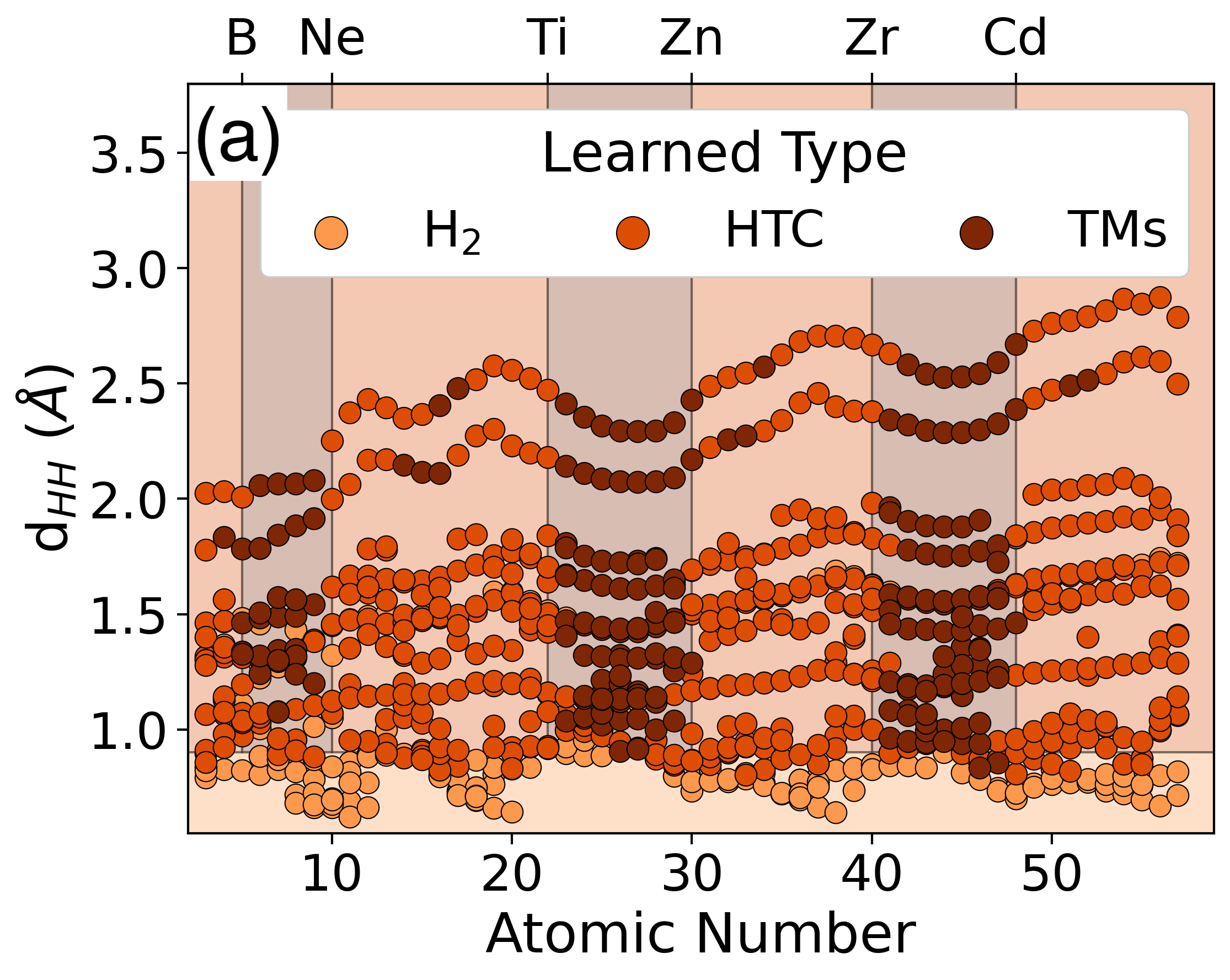}
\caption {Distribution of data points in DB1 with respect to minimum H-H distance \dHH~and atomic number $Z$. Dark brown, orange, and light orange dots correspond to cluster TMs, HTC and H$_2$, respectively.
\label{fig:CLUSTERING}}
\end{figure}

Fig.~\ref{fig:CLUSTERING} shows a scatter plot of the data in DB1 in the atomic number ($Z$) vs H-H distance \dHH, colored according to the assignment to one of three clusters: H$_2$, HTC, and TMs. 
We note that for clustering we used all the numerical features in DB1, none of which contains direct information on the chemical properties of the A atom, except for the atomic number.
Despite this, the clustering algorithm did find a partitioning which bears a striking similarity with the periodic table: the threshold between data points
classified in the HTC/TMs clusters coincides with the beginning of the first-row $p$-, and $d$-blocks of the periodic table. Upon closer inspection, we note that also a few
data points in the H$_2$ cluster have \dHH~$> 0.8$ ~\AA, indicating that clustering is not based on the sole \dHH~ feature.

Having established that the data can be separated into physically-meaningful clusters, which may also correlate with \tc{}, 
we are now interested in learning which of the features in our DB may be considered effective indicators (descriptors) of high-\tc{} behaviour.
 To this end, we performed a feature importance analysis after training a
 simple Random Forest Classifier \cite{breiman2001randomforests} to classify the data in DB2 into high- and low-\tc{}.
 
 After optimizing the model (see the Methods section for details) -- we obtained an average recall score of 0.79, a precision of 0.79, a $F1$ of 0.79, and an accuracy of 0.90. This result is comparable with similar studies when the dataset had a comparable size \cite{stanev2018mltc}, and would significantly improve with a larger dataset, 
 which could be easily obtained by systematically adding more templates.
 Given the relatively poor performance, we do not focus on the model itself, but only used it to obtain a qualitative identification of 
 the most relevant features. These are: \vol, \ADOS, \HDOS, BG, \BMq{}, Z, $\omega_{min}$, \rhoH, and \dH{}.
For most of these features, it is easy  to identify a posteriori a physical motivation for its correlation with high-\tc.
For two features, i.e. the Bader charge on the H atom (\rhoH) and the quasi-bulk modulus \BMq, this correlation
is less obvious. However, we will  show in the next section that both exhibit a strong variability across elements and prototypes,
and their distribution permits to draw very informative material maps of the phase space of high-pressure binary hydrides.

%


\begin{figure*}[!htb]
\centering
\includegraphics[width=1.00\columnwidth,angle=0]{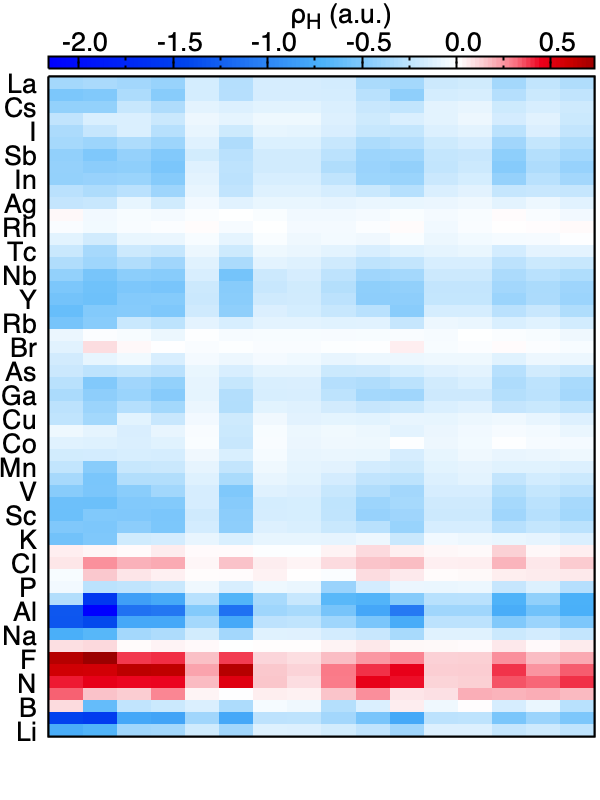}
\includegraphics[width=1.00\columnwidth,angle=0]{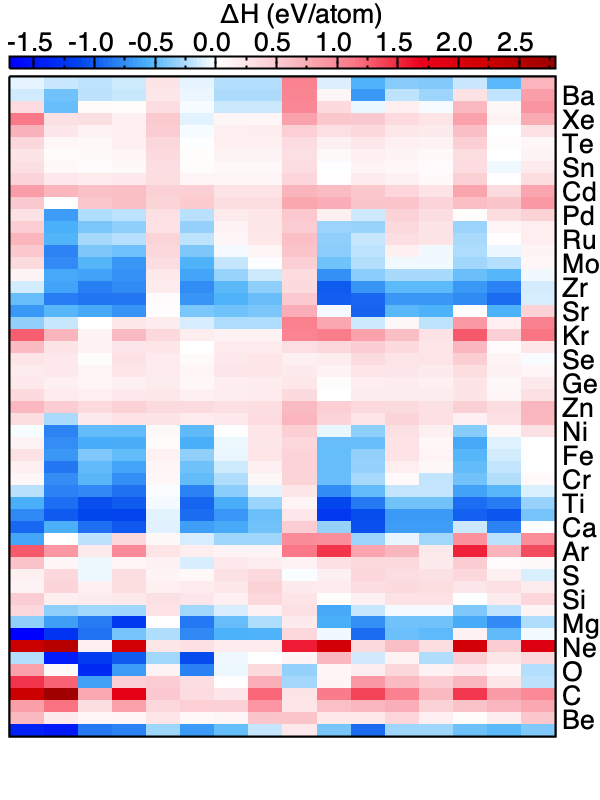}
\includegraphics[width=1.00\columnwidth,angle=0]{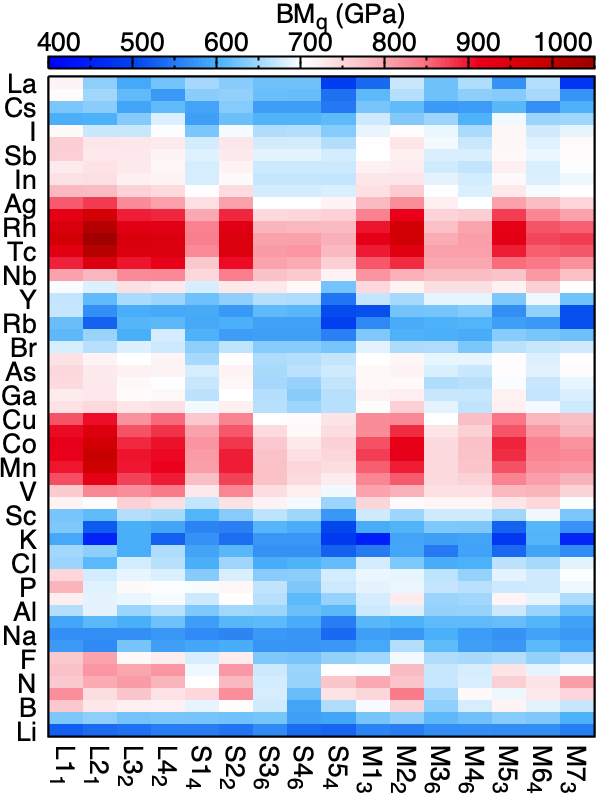}
\includegraphics[width=1.00\columnwidth,angle=0]{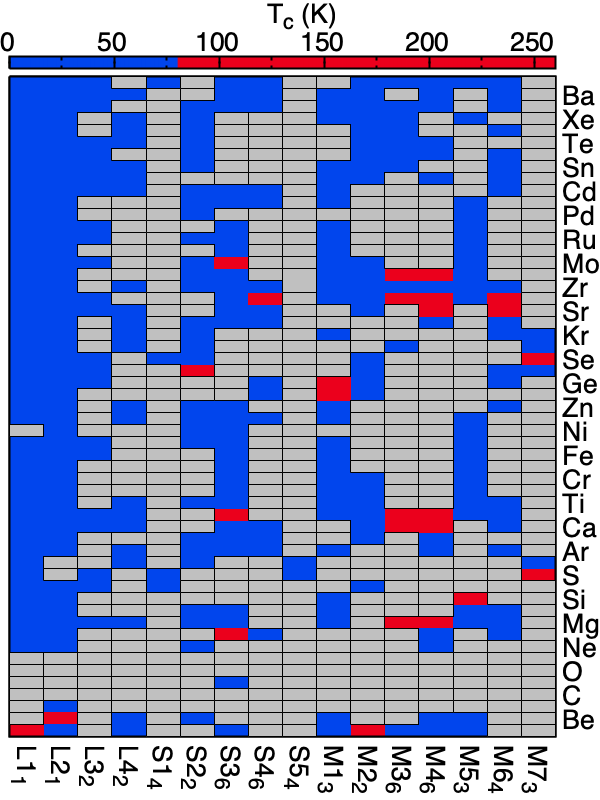}
\caption{
Heat map of different properties (color) of all the hydride structures 
for different elements (y-axis) and different templates(x-axis). All quantities
are estimated for pressure of 200 GPa.
(a) Average Bader charge on the H atom in a.u.; 
(b) Formation Enthalpy $\Delta $H (eV/atom) estimated w.r.t stable elemental phases;
(c) average quasi bulk-modulus (see SM) in GPa and (d) Critical superconducting
temperature \tc{} in $K$. Red represents high-\tc{} and blue low-\tc{}; the threshold is 
set to \tc{} = 77 $K$, corresponding to N$_2$ boiling temperature.
The methodology used to estimate different quantities are provided in the SM.}
\label{fig:HEATMAP}
\end{figure*}

\section{Structural Property Maps}
In this section we discuss some general trends that emerge from an in-depth analysis of DB1. 

In Fig.~\ref{fig:HEATMAP}, elements of the periodic table
and structural prototypes are arranged along the rows and columns
of a property matrix. The resulting boxes are colored with a blue to red heatmap (from lower to higher values), for four
different properties: from top to bottom and left to right, we show the average Bader charge on the H atom (\rhoH{}), the formation enthalpy (\dH), the quasi-bulk modulus (\BMq), and the critical temperature (\tc). Analogous heatmaps for
the other features of our DB1 are available in Fig. 1-8 in SM.
The choice of these four observables is motivated as follows: \dH{} gives information on the degree of chemical compatibility between the structure and the template, \rhoH{} describes the type of bonding, \BMq{} summarizes the lattice properties and quantifies the bonding strength and, as we will discuss later on, correlates with \tc{}. Last, \tc{} is the target property, which we want to compare with the other observables.

The formation enthalpy (\dH{}) heatmap describes the thermodynamic stability of different hydride templates across the periodic table: blue (red) corresponds to stable (unstable) hydrides. Rows with elements of the $p$-block and late transition metals (Si-Ar), (Cu-Kr) and
(Ag-Xe) are red to white, indicating that at 200 GPa the formation of binary hydrides with these elements is less likely. $s$-block elements (except Be) and early $d$-block elements form stable
binary hydrides (\dH~$\leq$ 0 eV/atom) across different templates.

These results are fairly consistent with available literature \cite{semenok2020distribution, shipley2021high, zurek2017alkalihydrides}, reporting that high-pressure hydrides  are mostly formed by electropositive elements w.r.t. H, as described by Pauling electronegativity\cite{pauling1932nature}. In Fig.~\ref{fig:HEATMAP}, electropositive elements exhibit a negative formation enthalpy for both S, M, and L-type templates, regardless of the hydrogen content. In other words, the same element can stabilize structures with different hydrogen content, hence with a different \dHH, chemical properties, and superconducting behavior. On the other hand, strongly electronegative elements, in particular first-row $p$-block elements, are only stable in L-type prototypes. Aside from first-row $p$-block elements, none of the other elements favors the formation of one specific template, while the opposite is true: a few specific templates, namely S1$_4$, S5$_4$ and M7$_3$ are strongly 
favored by specific A elements. For instance, the SH$_3$-like template (M7$_3$) is mostly unstable (\dH~$>$ 0), except for a few, selected elements (N, O, S, Sc, Ti). Indeed, in these structures the A-H interaction induces orbital configurations which are energetically favorable only for specific elements. 

The \rhoH{} heatmap in Fig.~\ref{fig:HEATMAP}(a) displays the average Bader charge on the H atom w.r.t. the neutral H atom for different combinations of host element A and template. 

The majority of the boxes are blue, indicating that H is generally more electronegative than most of the other elements, with the exception of late $d$-block elements (Co-Cu , Rh-Ag), halogens (Cl, Br, I), noble gases (Ar, Kr, Xe) and 2nd row $p$-block elements (B-Ne). The general behavior essentially follows an electronegativity trend analogous to the one at ambient pressure. However,
upon closer inspection, it can be noted that the columns where \dH{} switches from blue to red (and vice versa) correspond to Si, Ar, Ni, Kr, Pd, and Xe. 
These are precisely the elements where the calculated high-pressure electronegativity changes sign, according to a recently introduced high-pressure electronegativity scale   \cite{rahm2019electroneg_p, oganov2022ptable_hardness}. 

On top of these general trends, it is interesting to observe that in some cases the structural template also plays a crucial role. 
%
The S1$_4$, S3$_6$, S4$_6$, M3$_6$, M4$_6$ and
M6$_4$ templates exhibit a small spread in the \rhoH{} value which is close to 0, in stark contrast with the other templates. Among
the S-type templates, which contain H$_2$ molecules, \rhoH{} is close to 0 because of the charge neutrality of the hydrogen molecule \footnote{The S2$_2$ template is an exception, as H$_2$ molecules and isolated H atoms coexist, which is reflected in its Bader charge that differs from other S-like templates.}. On the contrary, among the M-type templates it is the weak-covalent nature of the hydrogen cage which suppresses the ionic exchange between hydrogen and the A atom.

In this dataset we observe that in most of the thermodynamically stable structures the A atom is an alkali metal, alkaline earth or early transition metal, and \rhoH{} is in the [ -0.5: 0.0 ] range. There are two exceptions to this trend: i) $p$-block elements, in which structures with positively charged hydrogen exhibit lower formation enthalpy, and ii) late transition metals, which present a value of \rhoH{} in the [ -0.5: 0.0 ] range, but are thermodynamically unstable, with a large positive \dH{}.

The heat map of the quasi-bulk modulus \BMq{} shown in Fig.~\ref{fig:HEATMAP}(c) summarizes the distribution of elastic properties. \BMq{} varies over a range from 400 to 1000 GPa. 
As a representative measure, the
well known SH$_3$, YH$_6$ and CaH$_6$ high-\tc{} superconducting phases have 698, 686, 628 GPa as \BMq{}, respectively. 

Similarly to \dH{}, \BMq{} depends more strongly on the A element than on the specific template. Elements of the $p$-, and $d$-block (Ti-Zn) and (Zr-Cd) exhibit \BMq{} larger than 700 GPa across different templates, while $s$-block elements and $(n-1)d^1$ elements have \BMq{} in the 500-700 GPa range, suggesting that the type of bonds formed in their binary hydrides are softer. 

The same templates for which \rhoH{} remains close to zero (S1$_4$, S3$_6$, S4$_6$, M3$_6$, M4$_6$ and M6$_4$) are an exception, as their \BMq{} is significantly lower than other templates with the same elements.  This occurs in S- and M-type templates, but never in L-type ones, where the bulk properties are dominated by the A element. In the following, we will refer to these templates collectively as \textit{soft} templates.

Panel (d) of  Fig.~\ref{fig:HEATMAP} divides compounds in DB2 into  low-\tc{} (blue) and high \tc{} (red) ones. Grey-coloured boxes indicate either non-metallic or metallic dynamically unstable structures. A significant portion of compounds in the \SHYDRA{} dataset hosts superconductivity (139 out of 880, i.e. 16\%), but only a relatively small fraction of those exhibits high \tc{}. (27, i.e. 19\%). 

Reassuringly, among the high-\tc{} structures we find well-known superhydrides: M7$_3$-SH$_3$ \cite{duan2014pressure, drozdov2015conventional}, M7$_3$-SeH$_3$ \cite{heil2015seh3sh3}, M3$_6$-MgH$_6$ \cite{feng2015mgh6}, M3$_6$-CaH$_6$ \cite{peng2017superconductivity, wang2012cah}, M3$_6$-YH$_6$ \cite{heil2019superconductivity, kong2021yh6}, S3$_6$-NaH$_6$ \cite{shipley2021high}. The same templates also exhibit a high-\tc{} with other A elements, such as M3$_6$-NbH$_6$ and ScH$_6$, and a few templates where high-\tc{} superconductivity has not been reported before exhibiting a surprising high-\tc{} behavior, such as M5$_3$-SiH$_3$, S2$_2$-AsH$_2$, M1$_3$-GeH$_3$. A complete list of the high-\tc{} structures is shown in Tab. \ref{tab:tc_templates_table}.

The majority of the high-\tc{} structures are formed by elements of the $s$-block, and M-type templates, where the formation enthalpy is negative. These are also regions for which the partial HDOS at the Fermi level is the largest (See Fig. 4 in SM). In particular, the M3$_6$ and M4$_6$ templates are the most favorable, hosting 5 and 6 high-\tc{} structures, respectively. In addition, although structures containing H$_2$ molecules are typically insulating, we do find a few high-\tc{} structures for templates with H$_2$ molecules: S3$_6$ (Na, Sc, Mo), S2$_2$ (As), and S4$_6$ (Y), as well as L1$_1$-LiH and L2$_1$-BeH. It is interesting to note that,  although the starting 
S2$_2$, S3$_6$ and S4$_6$  templates are S-type, the corresponding high-\tc{} structures formed with different A  elements listed in Tab. \ref{tab:tc_templates_table} 
are not classified in the H$_2$ cluster, with the exception of S3$_6$-NaH$_6$.
This indicates that unbiased classification based on unsupervised clustering has a clear advantage over human assumptions.

\begin{table}[htb]
\begin{tabular}{llllrrrr}
\toprule
  TEMPLATE & Cluster &  Formula &  HDOS &   $\lambda$ &   $\omega_{log}$ (K) &    T$_c$ (K) \\
\midrule
S2$_2$  &     HTC &  AsH$_2$ &     0.28   &  1.50 &   720     &   82    \\ 
M1$_3$  &     HTC &  GaH$_3$ &     0.69   &  0.96 &   1258    &   83    \\
M6$_4$  &     HTC &  YH$_4$ &      0.44   &  0.98 &   1289    &   87    \\
L2$_1$  &     HTC &  BeH$_1$ &     0.22   &  1.68 &   724    &   91    \\
L1$_1$  &     HTC &  LiH$_1$ &     0.86   &  8.21 &   386    &   91    \\
S3$_6$  &     H$_2$ &  NaH$_6$ &     0.90   &  2.95 &   516    &   92    \\ 
M3$_6$  &     TMs &  NbH$_6$ &     0.26   &  2.52 &   580    &   96    \\
M4$_6$  &     TMs &  NbH$_6$ &     0.26   &  2.52 &   581    &   96    \\
S3$_6$  &     HTC &  ScH$_6$ &     0.40   &  1.38 &   938    &   98    \\ 
S4$_6$  &     HTC &  YH$_6$ &     0.26   &  2.69 &   611    &   104    \\ 
M7$_3$  &     HTC &  SeH$_3$ &     0.47   &  1.02 &   1461    &   105    \\
S3$_6$  &     TMs &  MoH$_6$ &     0.48   &  2.20 &   763    &   117    \\ 
M1$_3$  &     HTC &  GeH$_3$ &     0.24   &  1.56 &   1037    &   123   \\
M6$_4$  &     HTC &  SrH$_4$ &     0.91   &  2.31 &   844    &   133    \\
M4$_6$  &     HTC &  SrH$_6$ &     0.94   &  1.42 &   1262    &   137    \\
M3$_6$  &     HTC &  ScH$_6$ &     0.33   &  1.93 &   1026    &   144    \\
M4$_6$  &     HTC &  ScH$_6$ &     0.33   &  1.92 &   1026    &   144    \\
M5$_3$  &     HTC &  SiH$_3$ &     0.38   &  1.41 &   1345    &   144    \\
M4$_6$  &     HTC &  YH$_6$ &     0.48   &  1.92 &   1323    &   185    \\
M3$_6$  &     HTC &  YH$_6$ &     0.48   &  1.92 &   1323    &   185    \\
M7$_3$  &     HTC &  SH$_3$ &     0.47   &  1.85 &   1371    &   186    \\
M4$_6$  &     HTC &  CaH$_6$ &     0.96   &  2.28 &   1259    &   197    \\
M3$_6$  &     HTC &  CaH$_6$ &     0.96   &  2.28 &   1259    &   197    \\
M2$_2$  &     HTC &  LiH$_3$ &     0.99   &  1.98 &   1406    &   201    \\
M3$_6$  &     HTC &  MgH$_6$ &     0.73   &  3.02 &   1260    &   228    \\
M4$_6$  &     HTC &  MgH$_6$ &     0.72   &  3.02 &   1260    &   228    \\
\bottomrule
\end{tabular}
\caption{High-\tc~binary hydrides in DB2 ($T_c \ge 77$ $K$).} 
\label{tab:tc_templates_table}
\end{table}

\begin{figure*}[htb]
\centering
\includegraphics[width=0.95\columnwidth,angle=0]{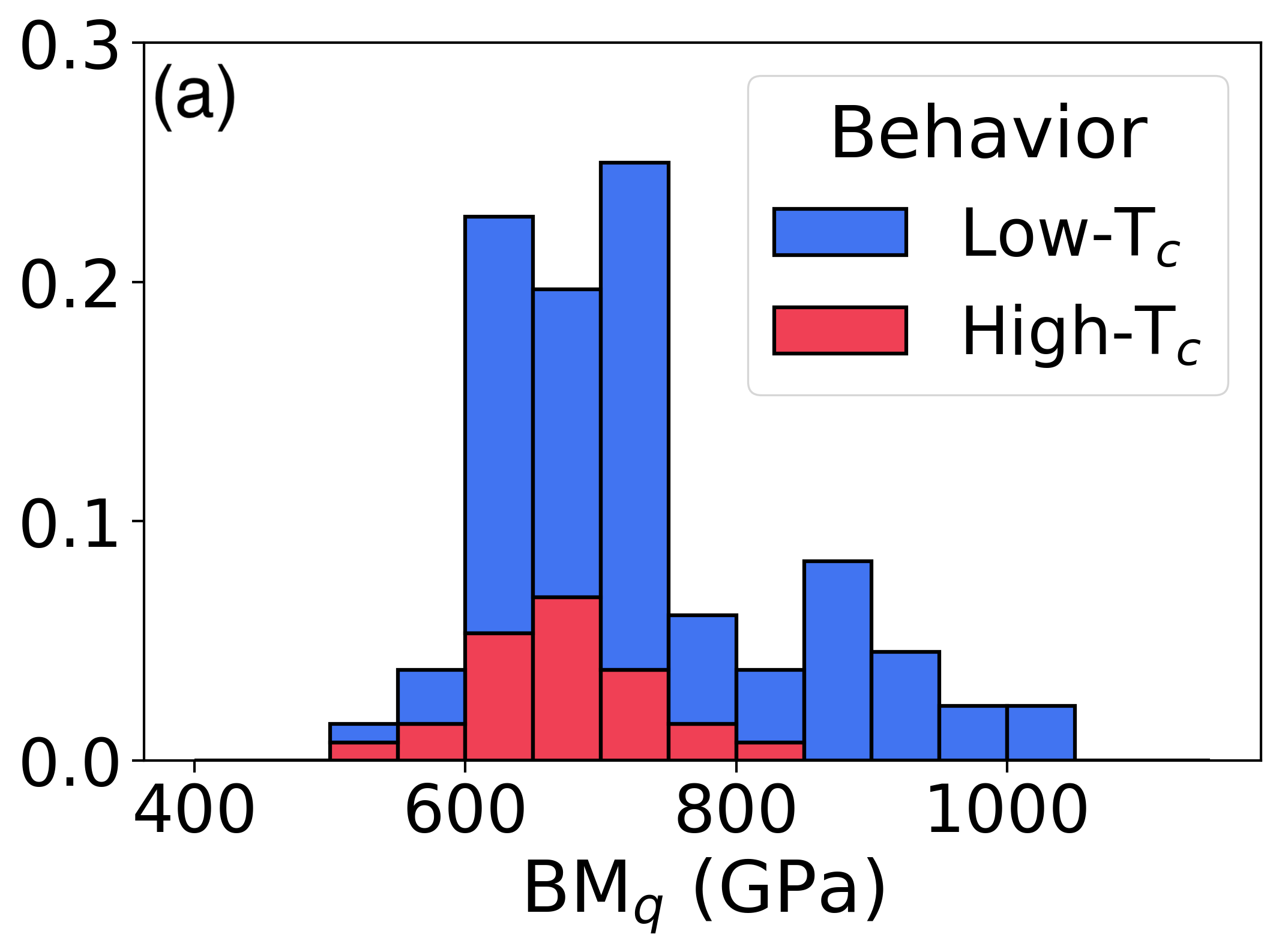}
\includegraphics[width=0.95\columnwidth,angle=0]{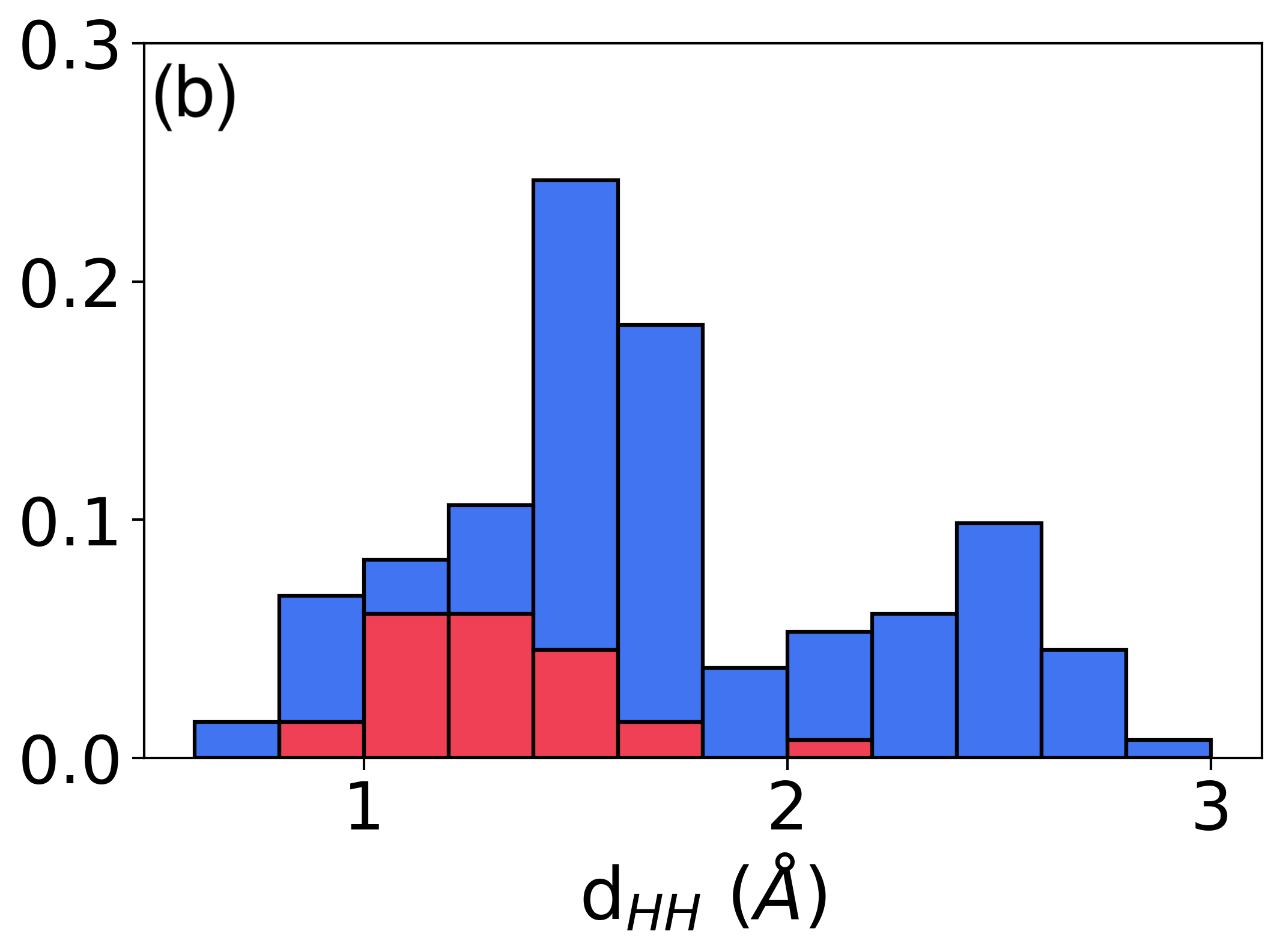}
\includegraphics[width=0.95\columnwidth,angle=0]{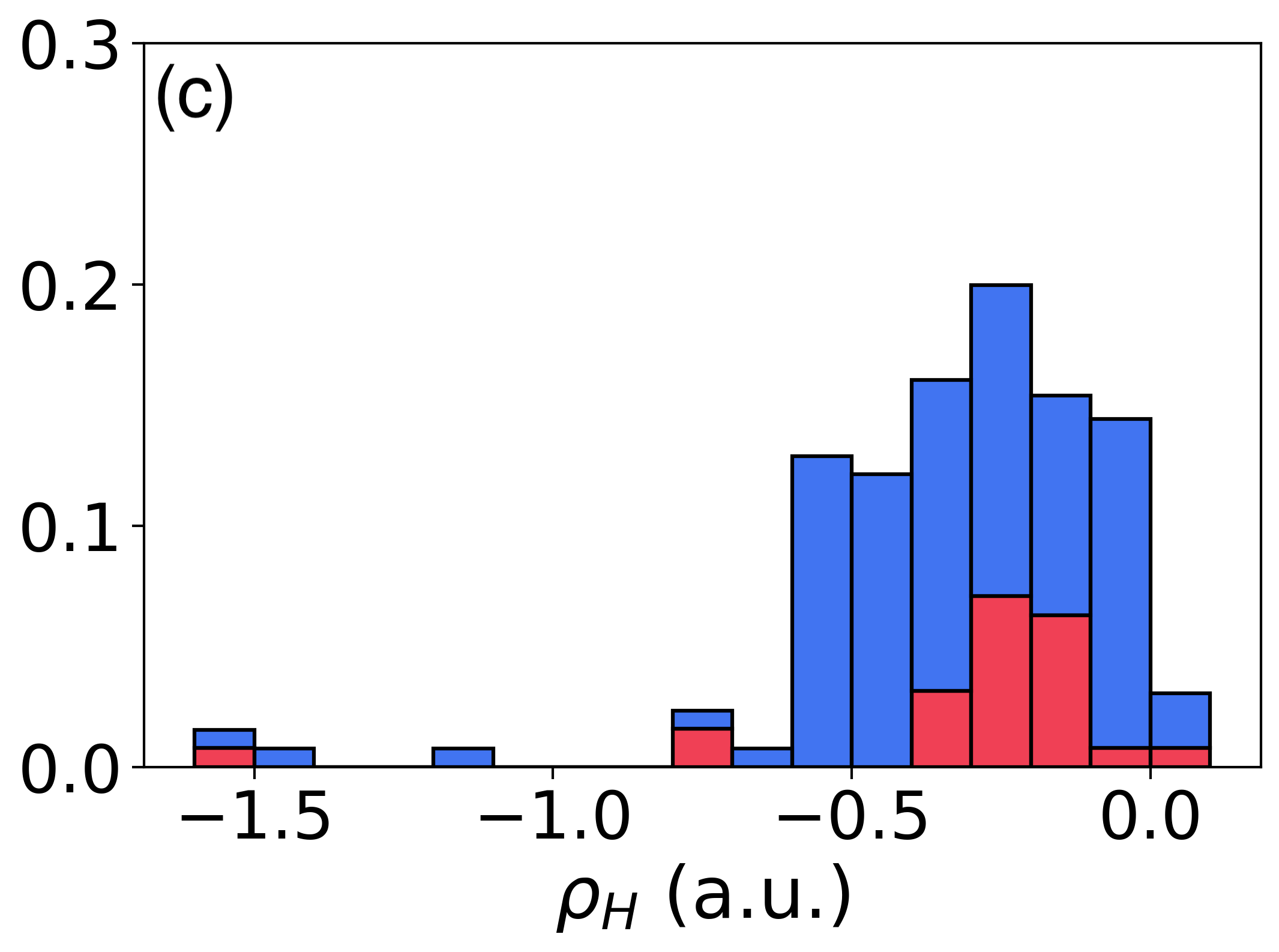}
\includegraphics[width=0.95\columnwidth,angle=0]{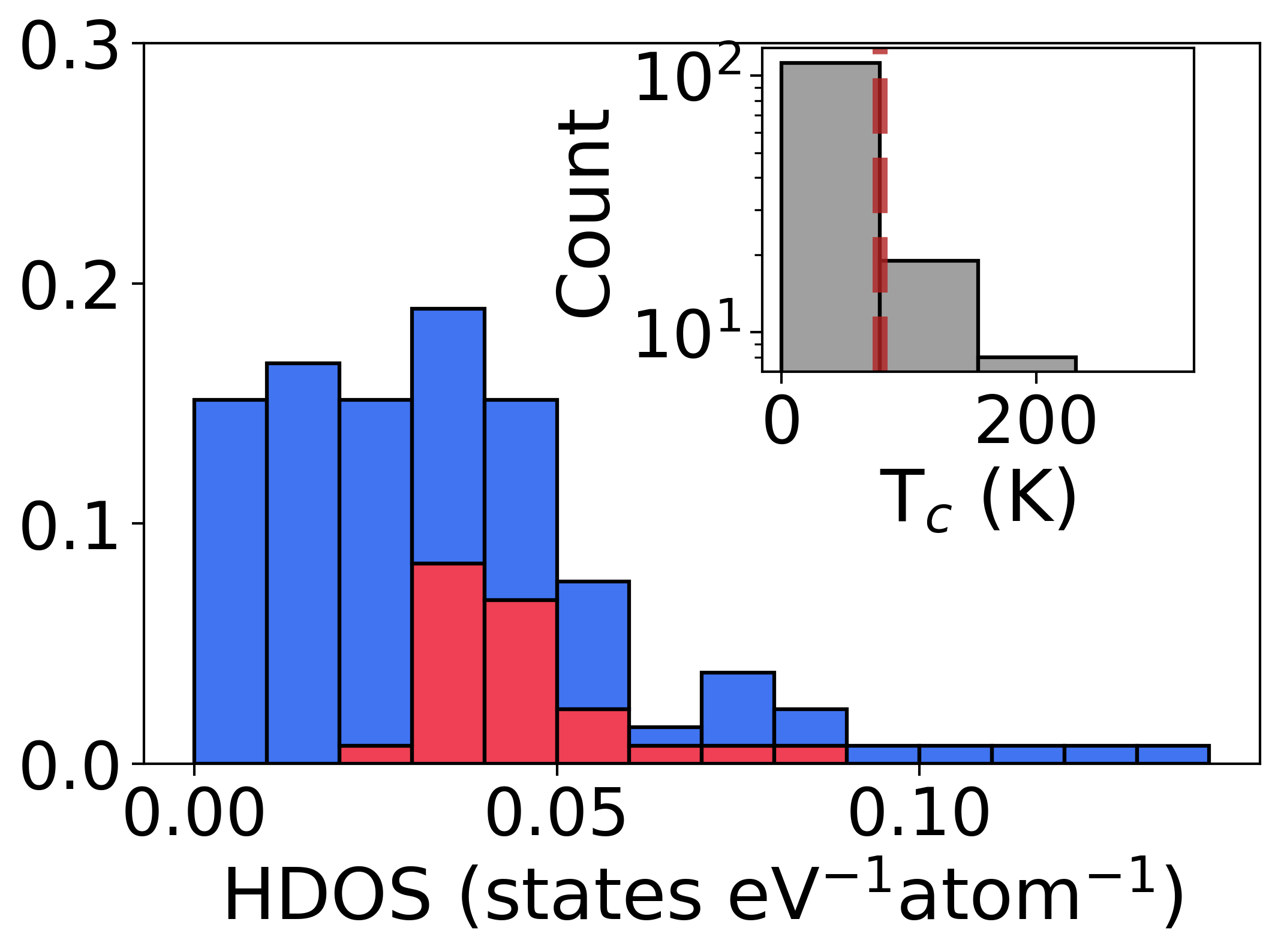}
\caption{Histograms of four quantities in DB2: quasi-bulk modulus \BMq, average nearest-neighbour H-H distance \dHH, average residual Bader charge on the H atom (\rhoH), and partial H DOS at the Fermi level HDOS (see text). Compounds are divided into low-\tc{} (blue) and high-\tc{} (red) structures. Inset: distribution of \tc{}'s in DB2 the red dashed line indicates the threshold for high-\tc{} superconductivity, given by liquid nitrogen boiling temperature (77 $K$).} 
\label{fig:histograms}
\end{figure*}

To better discuss the correlation between other observables and \tc{}, it is useful to look at aggregate data. In Fig.~\ref{fig:histograms} we show histograms with the distribution of \BMq, \dHH, \rhoH, and HDOS, broken down into low-\tc{} (blue), and high-\tc{} (red), with the same meaning as the previous figure. While low-\tc{} structures are uniformly distributed with respect to these four variables, it clearly emerges that high-\tc{} ones are concentrated in a narrow range of parameters. In particular:  500 < \BMq{} < 700 GPa, 0.8 < \dHH~ < 1.8 $\AA$, -0.4 < \rhoH{} < 0 a.u., and 0.03 < HDOS < 0.06 states eV$^{-1}$atom$^{-1}$. In other words, if one or more of these four physical observables are in the right range, this does not automatically imply a high-\tc{}, but in order to have high-\tc, the observables have to be in the right range (the conditions are necessary but not sufficient).
The main consequence is that, based on these observables, it would be possible to predict whether a structure can be a high-\tc{} superconductor with high sensitivity, but one must be willing to accept a significant fraction of false positives. Our result, however, confirms that all high-\tc{} hydride superconductors share similar features: 
a relatively soft H lattice, where H-H interactions are much weaker than in H$_2$ molecules, but still strong enough to ensure H-H bonds (\BMq, \dHH). This type of bonding becomes possible when enough negative charge is pumped into the hydrogen orbitals from other atoms acting as reservoirs (\rhoH) \cite{peng2017superconductivity, belli2021strong}. Moreover, the resulting electronic
structure should also ensure that electrons in H bonds are actively involved in the superconductivity pairing (HDOS).

\section{Conclusions}
In summary, in this paper we introduced a high-throughput approach for the computational generation of  databases of superconductors, 
and applied it to binary hydrides at high pressures.

Our MultiTemplate-HighThroughput approach
is based on the generation of template structures with crystal structure prediction methods,
carefully selected to be i) as chemically diverse as possible, and ii) high-symmetry. 
A database of structures is then generated by substituting elements from the whole periodic table in the non-hydrogen site
of all structural templates.

The resulting \SHYDRA{} database contains 880 potential hydride superconductors (DB1), characterized by various electronic, vibrational and chemical properties;
the 139 dynamically stable and metallic compounds form a database of superconductors (DB2), for which \tc{} was computed at the McMillan-Allen-Dynes level.
All data are available as a single CSV file in the SM.

An illustrative data-driven analyses  demonstrate that both DB1 and DB2 are information-rich and provide deep physical insight into the chemistry of high-\tc{} superhydrides. In particular, we find that four observables, namely the shortest H-H distance, the quasi-Bulk Modulus, the average residual Bader charge on hydrogen, and the hydrogen fraction of the DOS at the Fermi level (\dHH{}, \BMq{}, \rhoH{}, and HDOS{}, respectively) provide a set of necessary conditions for superconductivity. Based on unsupervised clustering, we propose a classification into three, physically-meaningful classes characterized by (1) short H-H bonds (H$_2$), (2) presence of $p$- and $d$-block elements (TMs) and (3) a soft lattice with \BMq{} lower than 800 GPa (HTC). The HTC cluster, in particular, hosts the 85\% of the predicted high-\tc{} superconductors.

We believe that the MTHT approach presents several advantages towards the generation of a database of conventional superconductors, compared to other approaches presented in literature\cite{shipley2021high,belli2021strong}. First, it is computationally efficient, as it avoids generating redundant data around ground-state structures that is typical of crystal structure prediction methods~\cite{rhorhofer2021features}. Second, it is easy to generalize by adding any number of templates to the basis. Third, it represents a much more complete exploration of the possible chemical interactions. Fourth, it effectively favours superconductors: 16\% of all generated structures are superconductors, and 3\% are high-\tc{}, about two orders of magnitude larger than one would have for structures generated with a structural search \cite{shipley2021high}.

Overall, the MTHT approach has proven to be a very effective method to map the phase space of binary hydrides, and could be generalized to the more complex space of ternary hydrides,  to provide much needed insight on their chemistry. As we have shown,  a careful choice of the initial templates is paramount to this end.
Extension to other classes of conventional superconductors is also straightforward, provided a more general definition of features. 
Our hope is that the \SHYDRA{} database will represent a much-needed first step of a generalized effort to map conventional superconductivity across the material space.

\section{Methods}
\subsection{First-principles Calculations}

First-principles calculations based on Density Functional Theory (DFT) 
was used for (i) structural predictions and (ii)  
multi-template high-throughput runs. Both types of runs were carried out using the 
Perdew-Burke-Ernzerhof (PBE) exchange-correlation(xc) functional~\cite{perdew1996generalized}, with an external pressure of 200 GPa.

For the structural prediction runs, we used the plane-wave based
Vienna ab-initio Simulation Package~\cite{VASP_Kresse}. 
The atoms were described by the inbuilt Projector Augmented Wave potentials~\cite{PAW_Bloch,kresse1999ultrasoft}. The geometric relaxations were carried out through multistep runs with increasing  kinetic energy cutoffs of 400 eV and 500 eV. Further details can be found in the SM.

For the high-throughput runs, we used the freely available plane-wave based \texttt{Quantum-Espresso}-6.5~\cite{giannozzi2017advanced} package; atoms were described by Optimized Norm Conserving Vanderbilt (ONCV) pseudopotentials~\cite{hamann2013optimized}, with a kinetic energy 
cutoff of 90 Ryd.  The Brillouin zone integration employed $\Gamma$-centered Monkhorst-Pack(MP) 8 $\times$ 8 $\times$ 8 electronic grids(\textbf{k}),
with a Methfessel-Paxton smearing of 0.04 Ry.
Further details are provided in the SM.

\subsection{Clustering}
Data analysis and clustering was performed using the pandas and keras python libraries. The number of clusters for the $k$-means algorithm was decided by examining the dendrogram shown in Fig. 10 in SM, constructed using agglomerative clustering with Ward linkage \cite{ward1963linkage}. This led us to our choice of $k$ = 3. The other possibly sensible choice ($k$ = 6) led to less physically meaningful results, and was discarded. The $k$-means clustering was then performed on the whole DB1. This algorithm partitions the data into $k$ clusters, each centered around a mean value, as to minimize the variance within each cluster. Conceptually, the algorithm is particularly suitable when the data can be divided into groups such that each group differs significantly from the others, while data points within each group are similar. 

\subsection{Classification}
To classify superconductors into high- and low-\tc{}, we used a Random forest (RF) classifier, as implemented in sklearn. RF is a powerful ensemble method based on simple decision trees. The choice of this algorithm was motivated by its rather good robustness against overfitting, and the need for a simple model, which could be trained sufficiently well on a little over than 100 data points. We employed a randomized 80/20 split in training and test set, with stratified sampling owing to the significant class imbalance. Statistical significance of the performance metrics was ensured by averaging over 100 iterations.

As a first step, we trained an initial model including all the features in our database, except those directly related to the critical temperature by Migdal-\'{E}liashberg theory (\tc{}, $\lambda$, $\omega_{log}$), and analyzed the importance of each feature, based on the Gini impurity decrease at each split, averaged over 2000 trees. The feature importance analysis is shown in Fig. 11 in SM. The final choice of features (See Section IV B.) was motivated by a combination of this analysis and a physical meaning of the features. In addition, we weighted the minority class (high-\tc{}) 20 to 1, as to compensate for the class imbalance, and improve the model recall at the expense of precision. 

As a second step, we optimized the hyperparameters using simple grid grid search. We found satisfactory results with a maximum tree depth of 5, a minimum number of samples per split of 4, and a number of estimators equal to 500, and a class weight of 20 to 1 between the high- and low-\tc{} class. This adjustment is intended to bias the model towards sensitivity to high-\tc{} structures, at the cost of precision, and compensates for the class imbalance of about 1 to 10 between high- and low-\tc{} structures.

\begin{acknowledgements}
S. Saha, S. Di Cataldo and W. von der Linden acknowledge computational resources from
the dCluster of the Graz University of Technology, the VSC3 of the Vienna University
of Technology, and support through the FWF, Austrian Science Fund, Project P 30269-N36
(Superhydra). S. Di Cataldo also acknowledges computational resources from CINECA, proj. IsC90\_HTS-TECH\_C. L. Boeri acknowledges support from Fondo Ateneo Sapienza 2018-20 and
computational Resources from CINECA, proj. Hi-TSEPH.
\end{acknowledgements}

\bibliographystyle{apsrev4-1}
\bibliography{main}

\end{document}